\documentclass[prl,aps,floatfix,superscriptaddress,footinbib,twocolumn,10pt,English]{revtex4}

\usepackage{amssymb,amsmath}
\usepackage{graphicx}
\usepackage{color}
\usepackage[colorlinks=true]{hyperref}
\usepackage{enumitem}
\usepackage{bm}

\newcommand{\grad}{\nabla}

\begin{document}  

\title{Tidal resonance in extreme mass-ratio inspirals} 

\author{B\'eatrice Bonga}
\affiliation{Perimeter Institute for Theoretical Physics, Waterloo, Ontario N2L 2Y5, Canada }
\author{Huan Yang}
\affiliation{Perimeter Institute for Theoretical Physics, Waterloo, Ontario N2L 2Y5, Canada }
\affiliation{University of Guelph, Guelph, Ontario N2L 3G1, Canada}
\author{Scott A.\ Hughes}
\affiliation{Department of Physics and Kavli Institute for Astrophysics and Space Research, Massachusetts Institute of Technology, Cambridge, Massachusetts 02139, USA}

\begin{abstract} 
%We show the existence of a new type of resonance in extreme mass-ratio inspirals (EMRIs): tidal resonance, induced by the tidal gravitational field of nearby stars or stellar-mass black holes.
We describe a new class of resonances for extreme mass-ratio inspirals (EMRIs): tidal resonances, induced by the tidal field of nearby stars or stellar-mass black holes.  A tidal resonance can be viewed as a general relativistic extension of the Kozai-Lidov resonances in Newtonian systems, and is distinct from the transient resonance already known for EMRI systems.  Tidal resonances will generically occur for EMRIs.  By probing their influence on the phase of an EMRI waveform, we can learn about the tidal environmental of the EMRI system, albeit at the cost of a more complicated waveform model.  Observations by LISA of EMRI systems therefore have the potential to provide  information about the distribution of stellar-mass objects near their host galactic-center black holes.
%This means that observations by LISA of EMRI systems can provide unprecedented information on the star and stellar mass distribution near the host supermassive black holes.
\end{abstract}

\maketitle 

{\noindent}{\bf Introduction}.~Ground-based gravitational-wave (GW) detectors have achieved tremendous success observing merging stellar-mass black holes (BHs) and neutron stars.  At lower frequencies ($\sim {\rm mHz}$), the Laser Interferometer Space Antenna (LISA) will probe binaries involving massive BHs at the centers of galaxies\cite{supermassivebh}.

One important source class for LISA are extreme mass-ratio inspirals (EMRIs), stellar-mass objects (typically a $10$--$30\,M_\odot$ BH) spiraling into a massive ($\sim10^5$--$10^7\,M_\odot$) BH in a galactic center. The large separation of mass scales means that the stellar-mass object's influence on the binary may be approximated as a perturbation of the large BH's spacetime. These stellar-mass objects  typically undergo $10^5$--$10^6$ orbits near the large BH in the LISA frequency band before finally plunging, providing a unique laboratory for mapping the spacetimes of BHs and enabling precise tests of strong-field gravity (see, for example, \cite{2019arXiv190303686B} for a recent review).
%The astrophysical population of these events heavily depends on the dynamical processes in galactic-center clusters, which is also related to the population of tidal disruption events (TDEs) that generate unique electromagnetic transients.

In this {\it Letter}, we propose that GW observations of EMRIs can be used to probe the environmental tidal field generated by stars and BHs near an EMRI system.  The EMRI waveforms will encode information about the BH and stellar distribution in galactic centers which are difficult to obtain with electromagnetic observations. We show that an environmental tidal field introduces a new type of resonance behavior, hereafter called {\it tidal resonance}, on the EMRI waveform. This effect can be intuitively understood as the general relativistic extension of the Newtonian Kozai-Lidov resonance \cite{naoz2016}.  Tidal resonances are different from transient resonances \cite{flanagan2012transient}, which arise from the gravitational self-force.
%and whose impact is small for low eccentricity orbits. 

\vspace{0.1cm}

{\noindent}{\bf BHs near EMRIs}.~Galactic centers are crowded environments. There are good theoretical reasons to expect several $10^5$ $M_\odot$ in stellar-mass BHs inside the inner parsec around a galaxy's central BH \cite{Morris:1993zz,Miralda_Escude_2000} and there is (tentative) observational evidence supporting this for our own galaxy \cite{Hailey2018ADC}.
Scattering processes can put stellar-mass objects (such as stars and black holes) near enough to the massive BHs in galactic centers for the object to be gravitationally bound to the BH.  
%Scatterings processes are thought to be important for producing tidal disruption events (TDEs), though there is currently an order of magnitude discrepancy between observed TDE rates and theoretical calculations, an important problem in galactic dynamics \cite{donley2002large,gezari2008uv,van2014measurement,magorrian1999rates,wang2004revised,stone2015rates}.  
Mean-motion resonance, in which a pair of stellar-mass objects jointly migrates towards the massive black hole until the resonant locking breaks down \cite{yang-li}, can also bring BHs close to the massive BH.

Currently, the distribution of stellar-mass objects nearby massive BHs is not well known. Proper dynamical theory calculations or $N$-body simulations are needed to compute the distribution of stellar-mass objects near galactic center BHs and assess the distance of the outliers closest to the central BH. 
Predictions based on a Fokker-Planck simulation suggest that a population of $40 M_\odot$ BHs can be  close  to Sagittarius A*, with median distance  $\sim 5 \; {\rm AU}$  \cite{emami2019segregation,emami2019gravitational}.   
This is roughly consistent with the following simple estimate for the distance of the closest BH, which mimics an argument in \cite{amaro2011butterfly}. The EMRI merger rate is about \cite{prospects}
\begin{align}\label{eqtau}
\frac{1}{T_{\rm EMRI}} \approx 0.3 \left ( \frac{M}{10^6 M_\odot}\right )^{0.19} {\rm Myr}^{-1}\,,
\end{align}
where $T_{\rm EMRI}$ is the interval between EMRI events, and $M$ is the mass of the central BH. Note that this estimate is subject to significant model uncertainties. Assuming that orbit decay is mainly driven by GW emission, at the time of an EMRI the distance $R$ to the next infalling BH (with mass $M_\star$) can be estimated using
\begin{align}
T_{\rm EMRI} \sim \frac{R}{\dot{R}} \sim \frac{5}{64} \frac{c^5 R^4}{G^3 M_\star M^2}\,,
\end{align}
telling us that
\begin{align}\label{eq:a}
R \sim 4.3\, \; {\rm AU} \left ( \frac{M_\star}{10 M_\odot} \right )^{1/4} \left ( \frac{M}{M_{\rm Sgr A^*}} \right )^{0.45} \,,
%a \sim 1.1\times10^{-5} {\rm pc} \left ( \frac{M_\star}{10 M_\odot} \right )^{1/4} \left ( \frac{M}{10^6 M_\odot} \right )^{0.45} \,.
\end{align}
with $M_{\rm Sgr A^*} = 4 \times 10^6 M_\odot$ the mass of Sagittarius A*.

Although it is interesting that this estimate agrees with \cite{emami2019segregation,emami2019gravitational}, we emphasize that it is only meant to provide a plausible case that a stellar mass black hole can be close enough for its tides to significantly influence an EMRI's orbital evolution.  In particular, this estimate ignores the fact that the tidal perturber's orbit will surely be eccentric. A critical point is that, because the tidal field scales as $M_\star/R^3$, the nearest outliers from a distribution of stellar mass black holes in the innermost regions of a galaxy (such as discussed in \cite{Morris:1993zz,Miralda_Escude_2000,Hailey2018ADC}) will have the strongest impact, significantly greater than the tides from another massive BH at $\sim 0.1 {\rm \; pc}$ (considered in \cite{yang2017general}).  The closest stellar-mass BHs are likely to be the main contributors to the tidal environment of EMRIs.

%%%%%%%%%%%%%%%%%%%%%%%%%%%%%%%%%%%%%%%%%%%%%%%%%%%%%%%%%%%%%%%%%%%%%%%%%%%%%%%%%%%%%%%%%%%
\vspace{0.1cm}
%%%%%%%%%%%%%%%%%%%%%%%%%%%%%%%%%%%%%%%%%%%%%%%%%%%%%%%%%%%%%%%%%%%%%%%%%%%%%%%%%%%%%%%%%%%
{\noindent}{\bf Tidal resonance}.~An EMRI orbit deviates from BH geodesic motion due to the gravitational self-force \cite{poisson2011motion} and the tidal field from nearby stars and BHs. The induced acceleration by the tidal field is generally smaller than that of the self-force. As we are interested in the EMRI motion near the central massive BH, it is natural to apply BH perturbation techniques \cite{poisson2011motion} instead of post-Newtonian simulations as was done in \cite{amaro2011butterfly}.

There is a two-timescale separation in the description of EMRI orbital evolution \cite{hinderer2008two}. This separation simplifies the analysis, approximating the orbit at any moment as a geodesic (with evolving integrals of motion) plus perturbations.  The fast timescale corresponds to the cyclic motion, and the slow timescale to the secular change of conserved quantities by radiation reaction. As Kerr geodesic motion  is separable \cite{misner2017gravitation}, it is convenient to use action-angle variables $q_{r,\theta,\phi}$ to describe the motion in $(r,\theta,\phi)$:
\begin{align}\label{eqexpansion}
\frac{d q_i}{d \tau} & = \omega_i ({\bf J}) +\epsilon \, g^{(1)}_{i,{\rm td}} (q_\phi, q_\theta, q_r, {\bf J}) + \eta \, g^{(1)}_{k, {\rm sf}}(q_\theta, q_r, {\bf J}) \nonumber \\
&+\mathcal{O}(\eta^2, \epsilon^2, \eta \epsilon)\,,\nonumber  \\
\frac{d J_i}{d \tau} & = \epsilon \, G^{(1)}_{i,{\rm td}} (q_\phi, q_\theta, q_r, {\bf J}) + \eta \, G^{(1)}_{i, {\rm sf}}(q_\theta, q_r, {\bf J}) \nonumber  \\
&+\mathcal{O}(\eta^2, \epsilon^2, \eta \epsilon)\,.
\end{align}
The action variables ${\bf J}:=\{J_r,J_\theta,J_\phi\}$ are functions of the energy $E$, angular momentum along the symmetry axis $L_z$ and the Carter constant $Q$; $\eta$ is the EMRI mass ratio, and $\epsilon := M_{\star} M^2/R^3$ characterizes the strength of the tidal field produced by the third body $M_\star$. The parameter $\tau$ is the proper time of the inspiraling body. The terms $G^{(1)}_i$ and $g^{(1)}_i$ force the orbit away from geodesic motion.  Terms with subscript ``td'' are from the tidal force, and depend upon the axial angle $\phi$ and the third body $M_\star$; terms with subscript ``sf" are from the self-force (generated by gravitational radiation reaction) and do not depend on $\phi$ and $M_\star$. Without the self-force and the tidal force, ${\bf J}$ would be conserved quantities and $q_i$ would increase at a fixed rate in time.

Focus now on the tidal force $G^{(1)}_{i,{\rm td}}$ and drop the subscript ``td.''  We write this term in the frequency domain
\begin{equation}\label{eq:fsum}
G^{(1)}_{i}(q_\phi, q_\theta, q_r, {\bf J}) =\!\!\sum_{m,k,n} G^{(1)}_{i, m k n}({\bf J}) e^{i (m q_\phi+k q_\theta +n q_r)}  \, ,
\end{equation}
with $m,k,n$ integer.
% \in \mathbb{Z}$.
Over the total duration of the EMRI inspiral ($\propto M/\eta$), the dissipative part of the self-force ($\propto \eta$) changes the conserved quantities by a fractional amount of order unity. In $G^{(1)}_{i}$, the exponential in $q_{\phi, \theta,r}$ generally oscillates in time, so a typical mode with nonzero $m,k,n$ will vanish after orbit averaging, and consequently does not contribute to secular changes of conserved quantities. However, in special cases one can have
\begin{align}\label{eqomega}
 \omega_{mkn} :=m \omega_{\phi}+k \omega_\theta +n \omega_r=0\,,
 \end{align}
so that the exponential does not oscillate.  If the corresponding force amplitude $G_{i,mkn}$ is non-zero, this mode will induce a secular change in ${\bf J}$. This is the tidal resonance.  By Eq.~\eqref{eqexpansion}, both ${\bf J}$ and $\omega_i(\bf J)$ change at the radiation reaction timescale $M/\eta$. The tidal resonance is thus transient because of the orbit's inspiral. However, it occurs under more general conditions than the transient resonance of the gravitational self-force \cite{flanagan2012transient}, which requires $k\omega_\theta + n \omega_{r} = 0$. Transient resonances have been shown to occur for generic EMRIs \cite{ruangsri2014census,brink2015orbital}; the same conclusion should apply for tidal resonances since its resonance condition is more general. Moreover, tidal resonances will exist for low eccentricity orbits, whereas the transient resonance may be unimportant for many LISA EMRI sources due to low eccentricity \cite{berry2016importance}.
%Furthermore, the tidal resonance is qualitatively different in Schwarzschild and Kerr spacetimes.  For Schwarzschild, tidal resonance occurs when $m \omega_\phi + n \omega_r=0$ (as the non-resonant motion is generically 2-dimensional rather than 3-dimensional) and the tidal force induces secular effects in the form of precession \cite{yang2017general}.
% I suggest leaving this out -- it's a fine point, but we don't explore it any further.  This might be a detail best explored in a longer analysis.

The tidal resonance induces a change in ${\bf J}$. Defining $\tau = 0$ as the moment of resonance, and expanding $q_i$ around this point as $q_{i 0} + \omega_{i 0} \tau +\dot{\omega}_{i 0} \tau^2 +\mathcal{O}(\tau^3)$, this change across the resonance is well-approximated by \cite{flanagan2012transient}
\begin{align}\label{eq:dj}
&\Delta J_{i}  = \epsilon \int^\infty_{-\infty}   G^{(1)}_{i} (q_\phi, q_\theta, q_r, {\bf J}) d \tau \\
&=\frac{\epsilon}{\eta^{1/2}} \sum_{s} \sqrt{\frac{2 \pi}{|\Gamma s|}} {\rm exp} \left [ {\rm sgn}(\Gamma s) \frac{i \pi}{4} +i s \chi\right ] G^{(1)}_{i, s m\, sk\, sn}\;,   \nonumber 
\end{align}
with $\chi :=m q_{\phi 0}+k q_{\theta 0}+n q_{r 0}$, $s$ a non-zero integer,
% \in \mathbb{Z}^{\neq}$
and $\Gamma := m \dot{\omega}_{\phi 0}+k \dot{\omega}_{\theta 0}+n \dot{\omega}_{r 0}$; terms with $s = \pm1$ dominate.  All quantities are evaluated at resonance. As $\Delta {\bf J}$ is proportional to $\epsilon/\eta^{1/2}$, the accumulated phase shift over $1/\eta$ inspiral cycles is proportional to $\epsilon/\eta^{3/2}$.

In Eq.~(\ref{eq:dj}), we ignore changes of the external tidal field during the resonance. This is valid if the orbital period of the perturbing third body, $T_{\rm td} \sim 2\pi\sqrt{R^3/M}$, is much longer than the resonance's duration, $T_{\rm res} \sim 1/\sqrt{\eta \Gamma}$.  When this holds, the tidal field is effectively static during the resonance.  It is possible that the third body is so close to the EMRI that $T_{\rm td} \lesssim T_{\rm res}$. In such a case, if the third body's orbit is near the EMRI's equatorial plane and has azimuthal frequency $\Omega_\phi$, we only need to correct $q_{\phi_0}$: the tidal resonance is shifted to $m (\omega_{\phi} \mp \Omega_\phi)+k \omega_\theta +n \omega_r=0$ (upper sign for prograde motion of the third body, lower for retrograde).  Because $\Omega_\phi \ll \omega_\phi$, such a resonance is dynamically the same as in the $T_{\rm td} \gg T_{\rm res}$ case, but is evaluated at a slightly different frequency.  In the most general setting, $G_{i}$ must include the motion of the third body or the time dependence of the tidal field in Eq.~\eqref{eq:dj}.

To evaluate $G_{i}$, we need the perturbation $h_{\alpha \beta}$ to the central BH's spacetime due to the tidal field.  This is found by solving Teukolsky's equation \cite{Teukolsky:1973} in the slow motion limit followed by metric reconstruction \cite{Yunes2006}.  For simplicity, we put the tidal perturber on the ($x$-$y$) equatorial plane and only consider its quadrupolar nature (the dipolar perturbations induced are zero), with the massive BH spin  along the $z$-axis \footnote{In general, the perturber's orbit should be eccentric. The tidal effect is larger when the perturber is close to the pericenter, and smaller when it is close to the apocenter. Multiple perturbers may contribute to the tidal environment. However, as the tidal strength sensitively depends on the distance, only closest BHs matter. }. As we will see, this restricts the type of resonances encountered.  Specifically, we choose as the tidal moment tensor $\mathcal{E}_{ab}=(M_\star/R^3) (2 \grad_a x \grad_b x - \grad_a y \grad_b y - \grad_a z \grad_b z)$, where $x$, $y$, and $z$ describe the motion of the perturbing third body in Cartesian-like coordinates (see Sec.~IX\,B of \cite{Poisson2004}).  We substitute this in Eqs.~(7), (45), and (46) of \cite{Yunes2006} to obtain $h_{\alpha \beta}$ in the ingoing radiation gauge in advanced Eddington-Finkelstein coordinates %
\footnote{
There is an overall factor of two missing in $h_{\alpha \beta}$ in \cite{Yunes2006} as $dL_z/dt$ at large radii yields half the Newtonian result. After correcting for this factor, $G^{(1)}_i$ agrees in the slow spin limit with $G^{(1)}_i$ for $h_{\alpha \beta}$ as in \cite{poisson2015tidal}.
}.
Next, we perform a coordinate transformation to Boyer-Lindquist coordinates.
Given $h_{\alpha \beta}$, we can compute the induced acceleration with respect to the background Kerr spacetime
\begin{align}\label{eq:acc}
a^\alpha & = -\frac{1}{2} (g^{\alpha\beta}_{\rm Kerr}+u^\alpha u^\beta)(2h_{\beta \lambda;\rho}-h_{\lambda \rho; \beta}) u^\lambda u^\rho\;,
\end{align}
with $u^\alpha$ the unit vector tangent to the worldline of the EMRI's small mass $\mu$.
The corresponding instantaneous change rates of the integrals of motion are \cite{yang2017general}
\begin{align}
	 \frac{d L_z}{d \tau} & = a_\phi  \\
	 \frac{d Q}{d \tau}& = 2 u_\theta a_\theta -2 a^2 \cos^2 \theta u_t a_t + 2 \cot^2 \theta u_\phi a_\phi \; .
\end{align}
The energy $E$ is conserved as the spacetime is assumed to be stationary during the resonance.

\begin{figure}
	\flushleft
	\includegraphics[width=8.8cm]{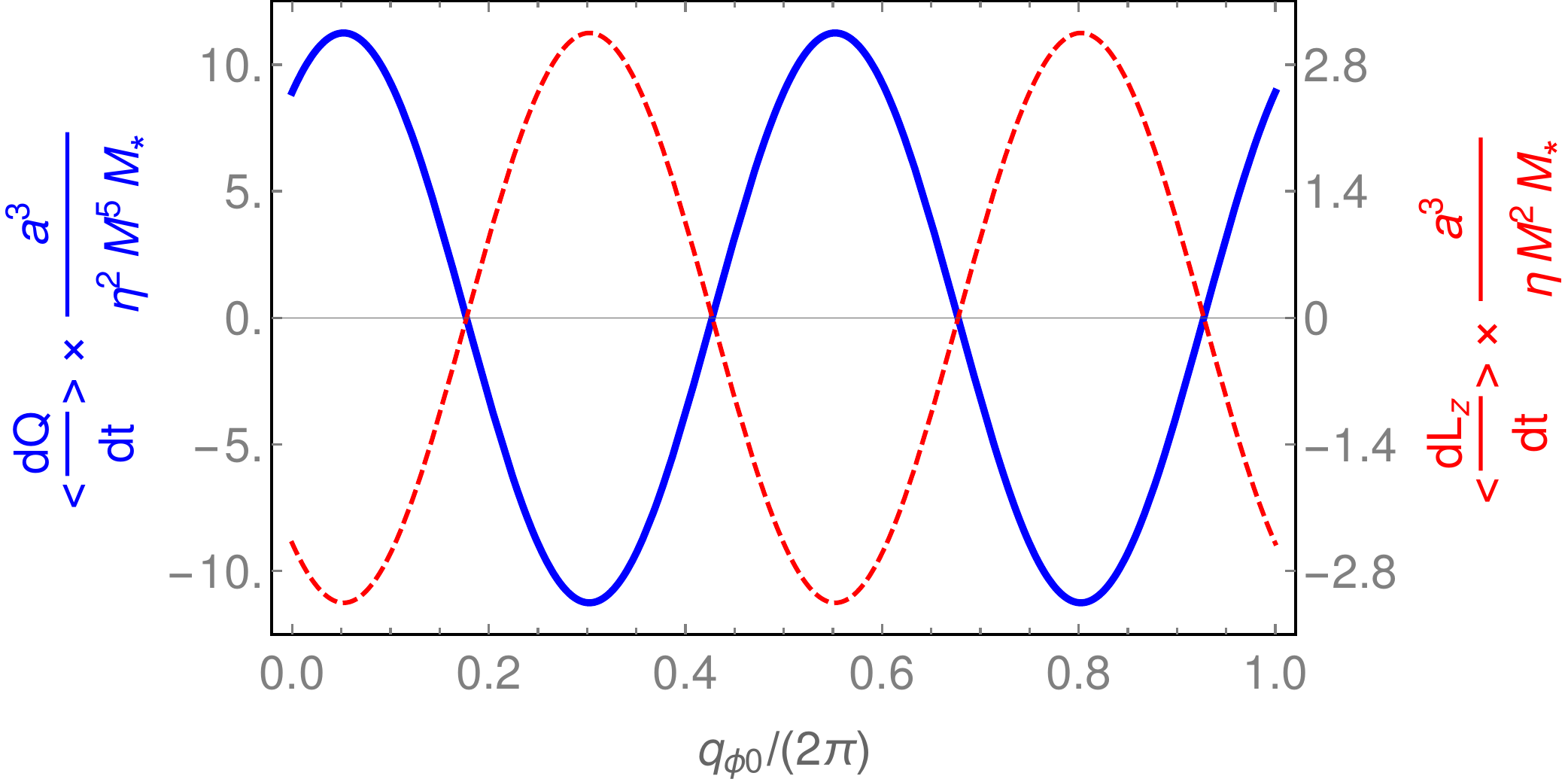}
	\caption{Average change rate of the Carter constant (solid, blue) and angular momentum along the z-direction (dashed, red) as a function of $q_{\phi0}$ for the case with $a=0.99$ (see Tab.~\ref{tab:results}).
	Both $\langle d Q/dt \rangle $ and $\langle d L_z/dt \rangle$ are normalized by $\epsilon$ to remove the associated linear dependence, and powers of $M$ to be dimensionless.
	}
	\label{fig:resql}
\end{figure}

%%%%%%%%%%%%%%%%%%%%%%%%%%%%%%%%%%%%%%%%%%%%%%%%%%%%%%%%%%%%%%%%%%%%%%%%%%%%%%%%%%%%%%%%%%%
\vspace{0.1cm}
%%%%%%%%%%%%%%%%%%%%%%%%%%%%%%%%%%%%%%%%%%%%%%%%%%%%%%%%%%%%%%%%%%%%%%%%%%%%%%%%%%%%%%%%%%%
{\noindent}{\bf Sample evolutions.} To illustrate the tidal resonance and to estimate its impact on the phase of an EMRI waveform, we consider three different scenarios summarized in Tab.~\ref{tab:results} and Fig.~\ref{fig:resql}.
In all these scenarios, the EMRI crosses a tidal resonance with $m: k : n = -2 : 2 : 1$ %
\footnote{
	The only non-zero resonances have $m=\pm2$ given that the tidal perturber is on the equatorial plane so that  $h_{\alpha \beta}$ only contains $m=\pm2$ modes (in principle, the metric should also include $m=0$ modes but those are not included in \cite{Yunes2006}).
	The $m: k : n = -2 : 1 : 2$ resonance vanishes because the tidal perturbation $h_{\alpha \beta}$ is reflection symmetric in the equatorial plane.
}.

\begin{table}[b]
	\caption{\label{tab:results}%
	Three prograde orbital motions. Fig.~\ref{fig:resql} shows the dependence on $q_{\phi0}$, which has the same functional form for all three cases. 
	}
	\begin{ruledtabular}
		\begin{tabular}{cccccc}
			\textrm{$a$\footnote{Dimensionless spin of the central BH.}}&
			\textrm{$r_{\rm min}$}&
			\textrm{$r_{\rm max}$}&
			\textrm{$\theta_{\rm min}$\footnote{$\theta_{\rm min} =\pi-\theta_{\rm max}$.}}&
			\textrm{$\dot{Q}_{-2,2,1}$} &
			\textrm{$\dot{L_z}_{-2,2,1}$} \\
			\colrule
			0.7 & 3.5 & 5.1628033  & $\pi/3 $ & $1.66 + 2.27i$ & $-0.35 - 0.47i$ \\
			0.9 & 3 & 6.6159726 & $\pi/4 $ & $6.60 + 7.70i$ & $-1.72 - 2.01i$ \\
			0.99 & 3 & 5.3718120 & $\pi/4 $ & $4.46 + 3.43i$ & $-1.23 - 0.95i$
		\end{tabular}
	\end{ruledtabular}
\end{table}

After orbit averaging, the sum in Eq.~(\ref{eq:fsum}) is 
%\footnote{
%	We verified explicitly that $G_{i, -2,2,1}$ (and its complex conjugate) is the only non-zero term on the right hand side; all other terms are zero to within numerical error.}
%
\begin{equation}
\left< G^{(1)}_{i}(q_\phi, q_\theta, q_r, {\bf J}) \right> \approx G^{(1)}_{i, -2,2,1}({\bf J}) e^{ -2 i  q_{\phi0}} + {\rm cc }  \, .
\end{equation}
With $G^{(1)}_{i,-2,2,1}$, we compute $\Delta Q, \Delta L_z$ as a function of $\chi$ using Eq.~\eqref{eq:dj}. For this, we also need $\Gamma$, which we calculate assuming that the main evolution of the orbit is due to GW dissipation. Within this approximation \cite{sago2005adiabatic,hirata2011resonant},
\begin{align}\label{eqjt}
&\left (\frac{\dot{J}_r}{\eta}, \frac{\dot{J}_\theta}{\eta}, \frac{\dot{J}_\phi}{\eta} \right ) \nonumber \\
&= - \sum_{l m k n} \frac{(n, k, m)}{2 \omega^3_{m k n}} \left ( |\tilde{Z}^{\rm out}_{l m k n}|^2+\alpha_{ l m k n} |\tilde{Z}^{\rm down}_{l m k n}|^2\right )\,,
\end{align}
where the coefficient $\alpha_{ l m k n}$, the asymptotic Teukolsky wave amplitude at infinity $\tilde{Z}^{\rm out}_{l m k n}$ and at the horizon $\tilde{Z}^{\rm down}_{l m k n}$ are defined in \cite{drasco2006gravitational} %
\footnote{The notation in \cite{drasco2006gravitational} is slightly different from that used here: $\tilde{Z}^{\rm out}_{lmkn}$ is denoted $Z^{\rm H}_{lmkn}$ in \cite{drasco2006gravitational}; $\tilde{Z}^{\rm down}_{lmkn}$ is denoted $Z^\infty_{lmkn}$.
}.  For a given resonance, we compute the wave amplitudes and $\alpha_{ l m k n}$ by solving the Teukolsky equation in the frequency domain, with a source term  associated with the stellar-mass object's orbital motion at frequencies $(\omega_r, \omega_\theta, \omega_\phi)$. Our code agrees very well with other Teukolsky equation solvers \cite{drasco2006gravitational}.

For the $a = 0.99$ initial conditions, $T_{\rm res} \sim ( \eta \Gamma)^{-1/2} \sim 14 \eta^{-1/2} M$ and the ratio between $T_{\rm res}$ and $T_{\rm td}$ is 
\begin{align}
\frac{T_{\rm res}}{T_{\rm td}} \sim 1.2 \; \left ( \frac{\mu}{10 M_\odot} \right )^{-\tfrac{1}{2}} \left ( \frac{M}{M_{\rm Sgr A^*}}\right )^{2} \left ( \frac{R}{4.3\, {\rm AU}}\right )^{-\tfrac{3}{2}}\,,
%\frac{T_{\rm res}}{T_{\rm td}} \sim 1.2 \; \left ( \frac{\eta}{2.5 \times 10^{-6}} \right )^{-\tfrac{1}{2}} \left ( \frac{M}{M_{\rm Sgr A^*}}\right )^{\tfrac{3}{2}} \left ( \frac{a}{4.3\, {\rm AU}}\right )^{-\tfrac{3}{2}}\,.
\end{align}
where $\mu$ is the mass of the small inspiraling body.  These timescales are comparable for this example, so we are in the regime $ T_{\rm res} \sim T_{\rm td}$ and must shift the resonance (including $\Omega_\phi$ in the resonance condition), as compared to the static perturber approximation. Since $\Omega_\phi/\omega_\phi \sim  7.1 \times 10^{-3}  (r/4 M_{\rm Sgr A^*})^{3/2}(R/4.3\, {\rm AU})^{-3/2}$,  this shift is negligible in evaluating the resonance strength.

\vspace{0.1cm}

\begin{figure}
\includegraphics[width=8cm]{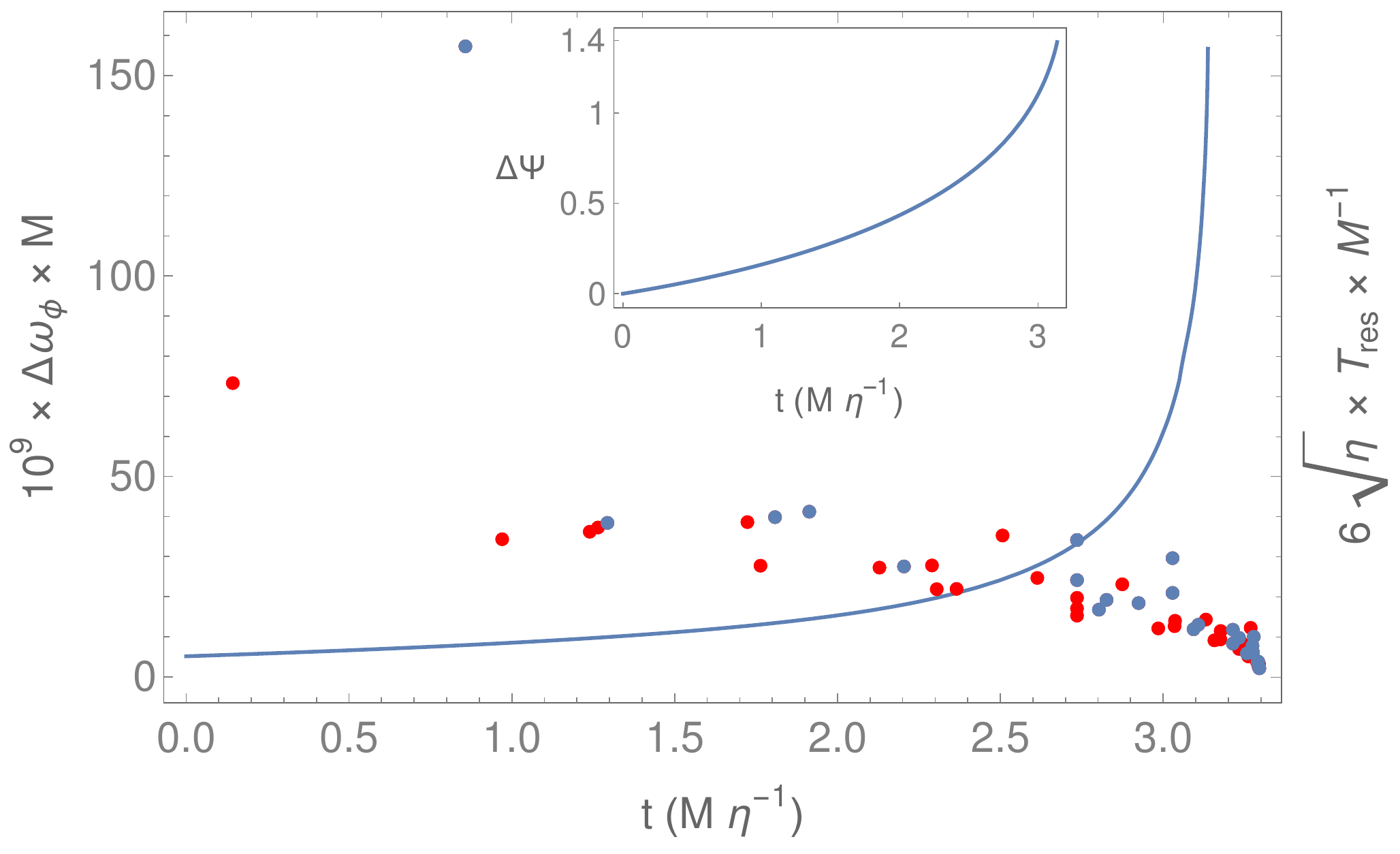}
\caption{
Evolution of the difference in $\omega_\phi$ between inspirals with and without resonant $\Delta J_{i}$ (blue curve), and illustration of resonances encountered during inspiral (dots).  We take the central black hole to have $M=M_{\rm Sgr A^*}$; both the inspiraling body $\mu$ and the perturbing tidal source $M_*$ are $10\,M_\odot$; and the tidal source is at separation $R=4.3\, {\rm AU}$.  The orbits start at the resonance point $q_{\phi 0}=0.33$ in Fig.~\ref{fig:resql}; the final time is the plunge. Red and blue dots show the resonance duration $T_{\rm res}$ for resonances with $\{|m|,|k|,|n|\} \le 5$; blue dots indicate $m=\pm 2$.  (The right-hand vertical axis has the same scale as the left.) The tightly bunched dots in the lower right illustrate how the system passes rapidly through multiple tidal resonances in quick succession as plunge is approached.
The inset shows the associated accumulated phase shift.
}
\label{fig:evo}
\end{figure}

{\noindent}{\bf Impact on orbital phase.}~To estimate the effect of tidal resonances on the phase of GW waveforms, we evolve two orbits starting at the point of tidal resonance considered in Fig.~\ref{fig:resql}, one with and one without $\Delta J_{i}$ included.  This evolution is realized with the orbit-averaged fluxes in Eq.~\eqref{eqjt} evaluated at each time step computed with the Teukolsky code, which in turn are used to update ${J_r, J_{\theta}, J_{\phi}}$ and subsequently $E, Q, L_z$ in time. At each time we compare $\omega_\phi$.  Its difference is plotted in Fig.~\ref{fig:evo}.  To estimate the deviation in orbital phase caused by the tidal resonance, we evaluate (c.f. Fig.~\ref{fig:evo})
\begin{align}\label{eq:psi}
&\Delta \Psi := \int^{T_{\rm plunge}}_0 2 \Delta \omega_{\phi} dt \nonumber \\
& = 1.4 \left( \frac{\mu}{10 M_\odot} \right)^{-\tfrac{1}{2}}\!\left ( \frac{M}{M_{\rm Sgr A^*}}\right )^{\tfrac{7}{2}}\!\left( \frac{M_*}{10\,M_\odot}\right)\!\left ( \frac{R}{{\rm 4.3\, AU}} \right )^{-3},
\end{align}
where $T_{\rm plunge}$ is the time of the plunge after the tidal resonance; in this example, $T_{\rm plunge} \simeq 0.78 (M/M_{\rm Sgr A^*})$ year. The factor of 2 in Eq.~\eqref{eq:psi} is because the strongest GW harmonic is the $m = 2$ mode.
%For simplicity, we have not included the contribution from the ringdown waveform, which may provide further phase deviation.
%Note from SAH: Ringdown is negligible for EMRI, not worth mentioning.
For systems with $R \lesssim 4.3\, {\rm AU}$ [as examined in Eq.~\eqref{eq:a}], the time till plunge is $\sim R^{4}$ [c.f.~Eq.~\eqref{eqtau}].  As such, the fraction of the population undergoing tidal resonances scales as $(R/4.3\, {\rm AU})^{4}$. However, it is important to note that the effect should be generally smaller for lighter massive BHs with less number of EMRI inspiral cycles.  

To estimate the phase resolution of EMRI measurements, we adopt the Fisher-information analysis presented in \cite{Lindblom2008,chatziioannou2017constructing}.  The statistical phase uncertainty roughly scales as $\sqrt{D-1}/{\rm SNR}$, where $D$ is the number of intrinsic source parameters in the waveform, and SNR is the measured signal-to-noise ratio.  By the Monte-Carlo study of \cite{gair2017prospects}, the number of EMRIs detected by LISA is likely to be $\mathcal{O}(10)-\mathcal{O}(10^3)$ per year at an SNR detection threshold of 20.  As SNR roughly scales as $1/d$ (with $d$ distance to Earth) and the number of sources per unit distance scales as $d^2$, we can estimate the average SNR of detected events to be $\sim30$. We thus roughly estimate the phase resolution to be $\Delta\Psi \sim 0.1$.  This suggests that the phase shift estimated in Eq.~(\ref{eq:psi}) should be easily detectable.  A significant fraction of EMRIs are likely to experience tidal resonances that induce $\Delta \Psi \ge 0.1$. Even if this holds for only $10\%$ of EMRI events, this corresponds to $\mathcal{O}(1)-\mathcal{O}(100)$ events per year.

The above estimate is based on a particular resonance for a single EMRI orbit.  A more rigorous calculation should survey a generic distribution of EMRI parameters and the mass/spin distribution of all host BHs.
% which will require a prescription for the spacetime of generic tidally perturbed Kerr BHs.  
It will also be important to include the influence of other signals which are simultaneously ``on'' during LISA observation, such as massive black hole inspirals, close white dwarf binaries in our galaxies, and other EMRI events which are being observed contemporaneously.  Most EMRI evolutions will cross multiple tidal resonances before plunge, as shown by the red dots in Fig.~\ref{fig:evo}. At early times, there are several resonances with duration comparable to the initial resonance which may contribute a comparable phase shift. Many short-lived tidal resonances cluster before the plunge due to the EMRI's rapidly changing orbital frequencies. Although their individual influence on the orbital phase is likely to be small compared to the initial resonance, there are many contributions.  These late resonances may also overlap, yielding collective effects.

\vspace{0.1cm}

{\noindent}{\bf Discussion}.~Similar to the Newtonian Kozai-Lidov effect, close orbits in a Kerr spacetime satisfying Eq.~(\ref{eqomega}) can be resonantly excited by an external tidal field, resulting in a secular shift in its orbital angular momentum \footnote{The orbit of the third body is generally averaged over when performing the analysis of Newtonian Kozai-Lidov effect, in which case $L_z$ of the inner orbit is also conserved, but the total angular momentum is not.}. EMRIs and tidal disruption events arise from the stellar clusters around massive BHs, and it has long been discussed that population studies of these events can be used to understand cluster properties and the growth history of massive black holes \cite{2019arXiv190303686B,pasham2019probing,stone2015rates,amaro2011impact}. Tidal resonance will enhance what is learned from EMRI events, providing additional data about other massive objects near galaxy centers, essentially probing the outliers of the stellar-mass distribution in these clusters.  
This information will come at the cost of a more complicated EMRI waveform model.  Much effort is currently going into making accurate self-force-based EMRI models, iterating in perturbation theory to second order in the mass ratio, and including effects like the impact of the smaller body's spin.  
Tidal resonances -- if not carefully modeled for -- may ultimately limit the precision to which it is worthwhile to make these waveform models. 
When testing General Relativity (GR) with EMRI observations in LISA, it is important not to miss attribute environmental effects as signals of GR violation.

% When unpredictable astrophysical systematics impact the phase at the several radian level, it may not be necessary (at least for measurement purposes) to make theoretical templates that are substantially more precise than this.

%The combined information from a population of tidal disruption events (TDE) observations and LISA EMRI measurements may provide more insight into the growth history of massive BHs and their impact on surrounding galactic stellar objects. For example, recent TDE measurement indicates that TDEs are more likely occurring in merging galaxies \cite{mattila2018dust}. It will be interesting to detect EMRIs in galaxies with measurable, strong tidal environments.

%So far we have only considered bound stars/BHs near the EMRI system. It is possible that during the resonances of an EMRI system within the LISA band, there are some fly-by stellar objects that approach closely to the EMRI to provide the tidal excitation. It will be interesting to analyze this effect and understand its observational imprints.
The Mathematica notebooks used for these calculations, including the metric perturbation and computation of $G_i$, are available upon request.
 
{\it Acknowledgements.} H.Y.\ thanks Christopher Hirata for sharing his Teukolsky code, and Eric Poisson for valuable discussions and comments.  
B.B.\ thanks Nicol\'as Yunes for sharing his Maple notebook with the tidally perturbed metric.  H.Y.\ acknowledges support from the Natural Sciences and Engineering Research Council of Canada. This research was supported in
part by the Perimeter Institute for Theoretical Physics.
Research at Perimeter Institute is supported in part by the
Government of Canada through the Department of
Innovation, Science and Economic Development Canada
and by the Province of Ontario through the Ministry of
Economic Development, Job Creation and Trade.  S.A.H.\ is supported by NSF Grant PHY-1707549 and NASA Grant 80NSSC18K1091.

%%%%%%%%%%%%%%%%%%%%%%%%%%%%%%%%%%%%%%%%%%%%
\bibliography{References}
\end{document}